\documentclass[aps,prb,twocolumn,superscriptaddress,unsortedaddress,amsmath,amssymb,showpacs,subeqn]{revtex4}
\usepackage{graphicx}
\usepackage{epsfig}
\usepackage{natbib}
\usepackage{color}

\begin{document}

\title{Time scales in the dynamics of an interacting
quantum dot}

\author{L.~Debora Contreras-Pulido}
\affiliation{Institut f\"ur Theorie der Statistischen Physik, RWTH
Aachen University, D-52056 Aachen, \& JARA - Future Information Technologies, Germany}

\author{Janine Splettstoesser}
\affiliation{Institut f\"ur Theorie der Statistischen Physik, RWTH
Aachen University, D-52056 Aachen, \& JARA - Future Information Technologies, Germany}

\author{Michele Governale}
\affiliation{School of Physical and Chemical Sciences and MacDiarmid
Institute for Advanced Materials and Nanotechnology, Victoria
University of Wellington, Wellington 6140, New
Zealand}

\author{J\"urgen K\"onig}
\affiliation{Theoretische Physik, Universit\"at Duisburg-Essen \&
CeNIDE, D-47048 Duisburg, Germany}

\author{Markus B\"uttiker}
\affiliation{D\'{e}partement de Physique Th\'{e}orique,
Universit\'{e} de Gen\`{e}ve, CH-1211 Gen\`{e}ve 4, Switzerland}

\date{\today}

\begin{abstract}
We analyze the dynamics of a single-level quantum dot with Coulomb interaction, weakly tunnel coupled  to an electronic reservoir,
after it has been brought out of equilibrium, e.g. by a step-pulse potential.
We investigate the exponential decay towards the equilibrium state, which is governed by three time scales. In addition to the charge and spin relaxation time there is a third time scale which is independent of the level position and the Coulomb interaction. This time scale emerges in the time evolution of physical quantities sensitive to two-particle processes.

\end{abstract}

\maketitle

\section{Introduction}
\begin{figure}[b]
   \includegraphics[width=0.9\linewidth,clip]{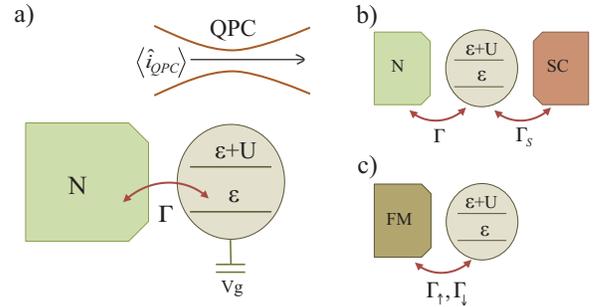}
   \caption{(Color online) Schematics of the models: a) Single level QD with Coulomb interaction, $U$, coupled to a normal lead with a tunneling strength $\Gamma$. Dot occupations can be measured via the current passing through a nearby quantum point contact (QPC) capacitively coupled to the dot. b) QD attached to an additional superconducting contact. c) QD coupled to a ferromagnetic lead.}
    \label{fig_scheme}
\end{figure}

The control and manipulation of single electrons in mesoscopic systems constitutes one of the key ingredients in nanoelectronics. The study of single-electron sources\cite{ebbecke04,feve07,blumenthal07,moskalets08,keeling08,kaestner08b,mahe10,battista11}  in the high-frequency regime has attracted a great interest due to their potential application in quantum electron optics experiments, in metrology, and in quantum information processing based on fermionic systems.\cite{beenakker05,samuelsson04,samuelsson05,feve08,splettstoesser09,mcneil11,hermelin11}
In this work we study the time evolution of a quantum dot (QD) tunnel coupled to a single electronic reservoir, as depicted schematically in Fig.~\ref{fig_scheme}(a).  In the presence of some time-dependent voltage modulations, this system defines the building block of the typical single-electron source, namely the mesoscopic capacitor.\cite{buttiker93} In the linear-response regime, the relaxation behavior of such a mesoscopic capacitor has been extensively studied theoretically\cite{nigg06,nigg08,ringel08,nigg09,mora10,hamamoto10,lopez11,filippone11,kashuba11} and experimentally,\cite{gabelli06} revealing the quantization of the charge relaxation resistance.\cite{buttiker93,nigg06,gabelli06,ringel08,mora10,hamamoto10,kashuba11}
On the other hand, the application of \textit{nonlinear} periodic potentials to the mesoscopic capacitor yields the controlled emission and absorption of electrons at giga-hertz frequencies.\cite{feve07,mahe10}
From these experiments the average charge as well as current correlations\cite{mahe10,albert10,partmentier11} after each cycle of the potential applied have been extracted. These results demonstrate the importance of investigating the dynamics of this kind of single-electron sources.   In some of the recent realizations\cite{gabelli06,feve07,mahe10} the Coulomb interaction is weak; however, in small-sized QDs the Coulomb blockade is, in general, strong and it is, therefore, desirable to include it in the theoretical analysis \cite{splettstoesser10a,hamamoto10,mora10,lopez11,filippone11,kashuba11} since it may even dominate time-dependent phenomena, see e.g. Ref.~\onlinecite{reckermann10}.
The time-evolution of interacting quantum dots after the coupling to the leads has been switched on, has, e.g., been studied in Refs.~\onlinecite{schoeller10,andergassen10,karrasch10,andergassen11}  and references therein.

Here we investigate the exponential relaxation of  a QD towards its equilibrium state after its has been brought out of equilibrium by applying, e.g., a voltage step pulse.
We consider a voltage pulse that affects the occupation of only a single
orbital energy level. The level can be spin split due to Coulomb
interaction.
In an earlier work,\cite{splettstoesser10a} some of the present authors investigated the decay of charge and spin of such a single level QD. It was found that the relaxation of charge and spin are given by rates which differ from each other due to Coulomb repulsion. Since the reduced density matrix of a QD with a single orbital level with spin is four dimensional, there are thus three rates which govern the relaxation of the diagonal elements of the density matrix towards equilibrium (plus one which is always zero and corresponds to the stable stationary state). In addition to the rates that govern charge and spin there is a third rate that appears in the relaxation of a single level QD with spin and with interaction. This additional rate is the subject of this paper.

Interestingly, this additional time scale is independent of the interaction and of the dot's level position. It is shown to be related to two-particle effects and appears, e.g., in the time-evolution of the mean squared deviations of the charge from its equilibrium value. We study in detail the perturbations leading to a relaxation of the system with the additional decay rate only, and find that it is indeed related to two-particle correlations. We also propose a procedure to separately read out the different relaxation rates occurring in the dynamics of the QD exploiting the sensitivity of a nearby quantum point contact to the occupation of the QD, see Fig.~\ref{fig_scheme} (a).

In order to further clarify the properties of the additional time scale, we extend our study to two other setups:  a QD proximized by an extra, superconducting electrode and  tunnel coupled to a normal lead; and a QD tunnel coupled to a ferromagnetic lead, see Fig.~\ref{fig_scheme} (b) and (c).

\section{Model}

We consider a quantum dot coupled to an electronic
reservoir.
We assume that the single-particle level spacing in the dot is
larger than all other energy scales, so that only one, spin-degenerate level of the QD spectrum is accessible. At a certain time $t_0$ the system is brought out of equilibrium, e.g. by applying a gate potential, and afterwards relaxes to an equilibrium dictated by the Hamiltonian $H=H_\mathrm{D}+H_\mathrm{T}+H_\mathrm{res}$. The Hamiltonian $H_\mathrm{D}$ of the decoupled dot
\begin{equation}
H_\mathrm{D}=\sum_{\sigma}\epsilon d_{\sigma}^{\dagger}d^{}_{\sigma}+U\hat{n}_{\uparrow}\hat{n}_{\downarrow}\ ,
\label{eq_hdot}
\end{equation}
contains the spin-degenerate level $\epsilon$ and the on-site Coulomb energy $U$ for double occupation of the dot. The creation (annihilation) operator of an electron with spin
 $\sigma=\uparrow,\downarrow$ on the dot is denoted by $d_{\sigma}^{\dagger}\left(d^{}_{\sigma}\right)$ and  $\hat{n}_{\sigma}$ is the corresponding number operator. The reservoir is modeled by the Hamiltonian
$H_\mathrm{res}=\sum_{k,\sigma}\epsilon_{k}c_{k\sigma}^{\dagger}c^{}_{k\sigma}$,
in which $c_{k\sigma}^{\dagger}\left(c^{}_{k\sigma}\right)$ creates
(annihilates) an electron with spin $\sigma$ and
momentum $k$ in the lead. The coupling between the dot and the reservoir is described by the tunneling Hamiltonian $H_\mathrm{T}=\sum_{k,\sigma}(Vc_{k\sigma}^{\dagger}d^{}_{\sigma}+\text{H.c.})$, where
$V$ is a tunneling amplitude, which we assume to be independent of momentum and spin. By considering a
constant density of states $\nu$ in the reservoir, the tunnel coupling strength $\Gamma$ is defined as
$\Gamma=2\pi\nu |V|^2$.

In the remainder of this paper, we focus on the relaxation behavior of the quantum dot to its equilibrium state and in particular on how this relaxation manifests itself in measurable quantities.
We are not interested in the dynamics of the reservoir, thus the
trace over its degrees of freedom is performed to obtain the reduced
density matrix of the QD. The Hilbert space is spanned by the four eigenstates of the decoupled dot Hamiltonian,
$ \{|\chi\rangle\}$, where $|0\rangle$ represents the unoccupied dot, the dot is in the state $|\sigma\rangle$ when being singly occupied  with spin $\sigma=\uparrow,\downarrow$, and $|d\rangle$ is the state of double
occupation. The energies related to these states are $E_0=0,E_\sigma=\epsilon$ and $E_\mathrm{d}=2\epsilon+U$, where we set the electrochemical potential of the reservoir to zero. As we consider spin-conserving tunneling events, the
off-diagonal elements of the reduced density matrix evolve independently of the diagonal ones (which are the occupation probabilities). We can, therefore, consider these probabilities alone, which arranged in a vector are given by $\mathbf{P}=(p_0,p_{\uparrow},p_{\downarrow},p_d)^\mathrm{T}$ and fulfill the condition $\sum_jp_j(t)=1$.  The time evolution of the occupation probabilities is governed by the generalized master equation
\begin{equation}
\frac{d\mathbf{P}(t)}{dt}=\int_{t_0}^{t}\mathbf{W}(t,t')\mathbf{P}(t')dt'\ ,
\label{eq_master}
\end{equation}
where the matrix elements $W_{\chi,\chi'}(t,t')$ of the kernel $\mathbf{W}(t,t')$ describe transitions from the state $|\chi'\rangle$ at time $t'$ to a state $|\chi\rangle$ at time $t$.

We consider now the dynamics of the system after being brought out of equilibrium at  time $t_0$.
Since for $t>t_0$ the total Hamiltonian is time independent, the transition matrix elements depend only on
the time difference $t-t'$, i.e. $\mathbf{W}(t,t')\rightarrow\mathbf{W}(t-t')$.
Furthermore, we are interested in the exponential decay towards equilibrium. To be more specific, we will therefore consider only the leading, time-independent, prefactor of the exponential functions. Time-dependent corrections to the pre-exponential functions, that generally may appear,\cite{schoeller10,buttiker00} are disregarded. Furthermore, when focussing on times $t$ distant from the switching time $t_0$, such that the difference  $t-t_0$ is hence much larger than the decay time of the kernel $\mathbf{W}(t-t')$, we can replace the lower limit of the integral in Eq.~(\ref{eq_master}) by $-\infty$.  Expanding the probability vector $\mathbf{P}(t')$ in Eq.~(\ref{eq_master}) around the measuring time $t$ we find\cite{splettstoesser10a}
\begin{equation}
\frac{d\mathbf{P}(t)}{dt}=\sum_{n=0}^\infty\frac{1}{n!} \partial^n{\mathbf{W}}\cdot\frac{d^n\mathbf{P}(t)}{dt^n}\ .
\label{eq_masterexpand}
\end{equation}
Here we introduced the Laplace transform of the kernel $\mathbf{W}(z)=\int_{-\infty}^{t}\mathbf{W}(t-t')e^{-z(t-t')}dt'$, with $\mathbf{W}=\left.\mathbf{W}(z)\right|_{z=0}$ and the $n$-th derivative of the kernel with respect to the Laplace variable  $\partial^n\mathbf{W}=\left[\partial^n\mathbf{W}(z)/\partial z^n\right]_{z=0}$. The formal solution of Eq.~(\ref{eq_masterexpand}) is given by
\begin{equation}
\mathbf{P}(t) = \exp({\mathbf{A}t})\mathbf{P}^\mathrm{in}\ ,
\label{eq_exp}
\end{equation}
which depends on the initial probability vector $\mathbf{P}^\mathrm{in}$ at $t=t_0$, where the initial values for the system parameters are given by the ones just after the switching time $t_0$.
The matrix $\mathbf{A}$ includes Markovian and non-Markovian processes.\cite{braggio06}
In the following, we consider the limit of weak coupling between quantum dot and reservoir and limit ourselves to a perturbation expansion up to second order in $\Gamma$, which is valid for the regime where the tunnel coupling $\Gamma$ is much smaller than the energy scale set by the temperature $k_\mathrm{B}T$. The perturbative expansion of $\mathbf{A}$ is $\mathbf{A}=\mathbf{A}^{(1)}+\mathbf{A}^{(2)}$ with $\mathbf{A}^{(1)}=\mathbf{W}^{(1)}$ and $\mathbf{A}^{(2)}=\mathbf{W}^{(2)}+\partial\mathbf{W}^{(1)}\cdot\mathbf{W}^{(1)}$, where the number in the superscript represents the power of $\Gamma$ included in the transition matrix $\mathbf{W}$. Notice that the first non-Markovian correction, i.e. the term $\partial\mathbf{W}^{(1)}\cdot\mathbf{W}^{(1)}$ is present in second-order in the tunnel coupling.
The evaluation of the kernel within a perturbative expansion can be performed using a real-time diagrammatic technique,\cite{konig96a,konig96b} which has been used in Ref.~\onlinecite{splettstoesser10a} in order to extract the exponential decay of spin and charge in the system studied here.
Considering Eq.~(\ref{eq_exp}), we see that the rates defining the decay of the state into equilibrium are found from  the eigenvalues of the matrix $\mathbf{A}$, which turn out to be real and non-positive.  The matrix $\mathbf{A}$ is not Hermitian, as expected since we deal with a dissipative system, and hence has different left and right
eigenvectors, $\mathbf{l}_i$ and $\mathbf{r}_i$.

\section{Results}

\subsection{Relaxation of the reduced density matrix}

 The time-dependent probability vector, $\mathbf{P}(t)$, can be expressed in terms of the  right eigenvectors of  $\mathbf{A}$, each being related to a decay with a different rate.
 The left eigenvectors determine the observable that decay with a single time scale only, see also the appendix. \\
 In the following we discuss the  exponential relaxation towards equilibrium of the vector of occupation probabilities, in first order in the tunneling strength $\Gamma$.
\subsubsection{Noninteracting limit}

We start by briefly discussing the simplest case of a single spinless particle. This limit is obtained, when a magnetic field much larger than the temperature is applied, $B\gg k_\mathrm{B}T$. The Hilbert space of the system is two dimensional and spanned by the states $|0\rangle$ and $|1\rangle$ for the empty and singly-occupied dot respectively, whose occupation probabilities are arranged in the vector $\mathbf{P}=(p_0,p_1)^T$.  The decay to the stationary state is governed by matrix $\tilde{\mathbf{A}}^{(1)}$ (defined equivalently to $\mathbf{A}^{(1)}$ but for the two-dimensional Hilbert space for the problem at hand) which contains a single relaxation rate, namely the tunnel coupling $\Gamma$, as intuitively expected.

We now include the spin degree of freedom but disregard interactions. The system is described by two independent Hilbert spaces spanned by the states $|0_{\sigma}\rangle$ and $|1_{\sigma}\rangle$ with $\sigma=\uparrow,\downarrow$.
The probability vector for each spin $\sigma$ can be written in terms of the eigenvalues and eigenvectors of the matrix $\tilde{\mathbf{A}}^{(1)}$ (for the two-dimensional Hilbert space) as
\begin{equation}
\mathbf{P}_{\sigma}(t)=\mathbf{P}^\mathrm{eq}_{\sigma}+e^{-\gamma_{\sigma} t}\left(\begin{array}{c}
1\\
-1
\end{array}\right)\left[\langle \hat{n}_{\sigma}\rangle^{\mathrm{eq}}-\langle \hat{n}_{\sigma}\rangle^{\mathrm{in}}\right]
\label{eq_psigma}
\end{equation}
where the right eigenvector corresponding to the eigenvalue zero of $\tilde{\mathbf{A}}^{(1)}$ defines the occupation probabilities for the equilibrium state, $\mathbf{P}^{\text{eq}}_{\sigma}=(p_{0,\sigma}^{\mathrm{eq}},p_{1,\sigma}^{\mathrm{eq}})^\mathrm{T}=(1-f(\epsilon),f(\epsilon))^\mathrm{T}$, with the Fermi function $f(\epsilon)=[1+\exp(\beta\epsilon)]^{-1}$ and the inverse temperature $\beta=1/k_\mathrm{B} T$.
Furthermore, $\hat{n}_{\sigma}=(0,1)$ is the vector representation of the number operator for dot electrons with spin $\sigma$, whose initial/equilibrium expectation value is obtained by multiplying it from the left into the initial/equilibrium probability vector, $\langle\hat{n}_{\sigma}\rangle^{\mathrm{in/eq}}=\hat{n}_{\sigma}\cdot\mathbf{P}^\mathrm{in/eq}_{\sigma}$. The rate $\gamma_{\sigma}=\Gamma$ is obtained as the negative of the non-zero eigenvalue of $\tilde{\mathbf{A}}^{(1)}$, with the corresponding left eigenvector being $\mathbf{l}_{\sigma}=(0,1)-\langle\hat{n}_{\sigma}\rangle^{\text{eq}}(1,1)$.

The time evolution of the occupation of each spin state is governed by a single decay rate $\Gamma$,
\begin{equation}
\langle \hat{n}_{\sigma}\rangle(t)=\langle \hat{n}_{\sigma}\rangle^{\mathrm{eq}}\left(1-e^{-\Gamma t}\right)+\langle \hat{n}_{\sigma}\rangle^{\mathrm{in}}e^{-\Gamma t}.\label{nsigma}
\end{equation}
This equation can be obtained making use of  the fact that the time evolution of the expectation value of any operator,
which describes
an observable of the QD, is given by projecting its vector representation from the left onto Eq. (\ref{eq_psigma}).

The time evolution of the total charge of the dot, $\langle \hat{n}\rangle(t)=\langle \hat{n}_{\uparrow}\rangle(t)+\langle \hat{n}_{\downarrow}\rangle(t)$, is also determined by a single relaxation rate $\gamma_{\sigma}=\Gamma$. This means that both charge and spin, which are quantities related with single-particle processes, do not evolve independently from each other and the corresponding decay is given by the same rate. A similar non-interacting problem has been studied \textit{non-pertubatively} in  Refs. \onlinecite{moskalets08} and \onlinecite{battista11}.

As a next step we consider the squared deviation of the charge from its equilibrium value, $[\hat{n}-\langle \hat{n}\rangle^{\text{eq}}]^2$. Its time evolution is obtained from Eq.~(\ref{eq_psigma}) as
\begin{eqnarray}
&&\langle[\hat{n}-\langle \hat{n}\rangle^{\text{eq}}]^2\rangle(t)-[\langle\hat{n}\rangle^{\text{eq}}]^2\\
&&=\sum_{\sigma=\uparrow,\downarrow}[1+\langle\hat{n}_{\sigma}\rangle^{\text{eq}}]\langle\hat{n}_{\sigma}\rangle(t)+2\langle\hat{n}_{\uparrow}\hat{n}_{\downarrow}\rangle(t)\nonumber
\end{eqnarray}
The last, two-particle term of this expression exhibits a decay rate given by $\exp(-2\Gamma t)$. This is in contrast to the spinless case, where such a term does not appear since double occupation is not possible.

Such an additional exponential decay with the rate $2\Gamma$ appears directly in the time evolution of the probability vector, when considering the full two-particle Hilbert space spanned by the basis $\{|0\rangle, |\uparrow\rangle, |\downarrow\rangle, |d\rangle\}$. In this basis,  Eq.~(\ref{eq_exp}) for the non-interacting regime can be written as:

\begin{eqnarray}
\label{ptot}
\mathbf{P}(t)  =  \mathbf{P}^\mathrm{eq}+\left(\begin{array}{c}
-\left[1-f(\epsilon)\right]\\
\frac{1}{2}\left[1-2f(\epsilon)\right]\\
\frac{1}{2}\left[1-2f(\epsilon)\right]\\
f(\epsilon)
\end{array}\right)
e^{-\Gamma t}\left(\langle\hat{n}\rangle^\mathrm{in}-\langle\hat{n}\rangle^\mathrm{eq}\right)\nonumber \\
+\left(\begin{array}{c}
0\\
\frac{1}{2}\\
-\frac{1}{2}\\
0
\end{array}\right)
e^{-\Gamma t}\langle\hat{s}\rangle^\mathrm{in}
+\left(
\begin{array}{c}
-1\\1\\1\\-1
\end{array}
\right)
e^{-2\Gamma t}\left(
\langle\hat{m}\rangle^\mathrm{in}-\langle\hat{m}\rangle^\mathrm{eq}
\right) \nonumber\\
\end{eqnarray}
where as before, $\mathbf{P}^{\mathrm{eq}}$ defines the state at equilibrium. The decaying part of the probability vector can be divided into three contributions which appear depending on how the initial state at $t_0$ differs from the equilibrium state. Deviations of charge and spin from their equilibrium value relax with the same rate $\Gamma$. The corresponding expectation values are calculated by multiplying the probability vector Eq.~(\ref{ptot}) from the left with the vector representation of the operators $\hat{n}=(0,1,1,2)$ and $\hat{s}=(0,1,-1,0)$ which represent the charge and spin, respectively, in this two-particle basis. The two left eigenvectors of the matrix $\mathbf{A}^{\mathrm{(1)}}$ with the same eigenvalue $-\Gamma$, are given by $\mathbf{l}_n=\hat{n}-\langle\hat{n}\rangle^{\mathrm{eq}}(1,1,1,1)$ and  $\mathbf{l}_s=\hat{s}$.

The third contribution to the decay of the system into the equilibrium comes from the relaxation rate $2\Gamma$, which enters the probability vector in connection with a quantity $\hat{m}$, defined by the operator in vector notation
\begin{equation}
\hat{m}=\left(
0,f(\epsilon),f(\epsilon),-1+2f(\epsilon)
\right).
\label{m}
\end{equation}
The left eigenvector of $\mathbf{A}^{(1)}$ with the eigenvalue $-2\Gamma$ is given by $\hat{m}-\langle\hat{m}\rangle^\mathrm{eq}(1,1,1,1)$. In contrast to charge and spin, the quantity represented by $\hat{m}$ does not have a straightforward intuitive interpretation, since it depends on the quantum dot parameters at $t>t_0$ and on the temperature and chemical potential of the reservoir via the Fermi functions.

\subsubsection{Finite Coulomb interaction $U$}

\begin{figure}[h]
   \includegraphics[width=0.91\linewidth,clip]{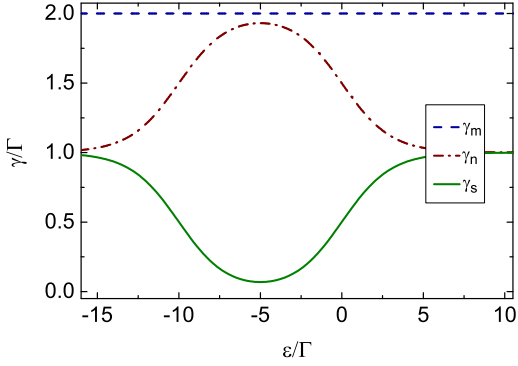}
   \caption{(Color online) Decay rates $\gamma_m$ (blue, dashed line), $\gamma_n$ (red, dash-dotted line) and $\gamma_s$ (green, solid line)  in units of $\Gamma$ as a function of the dot level position $\epsilon$. The temperature is $k_\mathrm{B}T=1.5\Gamma$ and the interaction energy is $U=10\Gamma$.  }
    \label{fig_decay}
\end{figure}

From now on we assume a finite on-site Coulomb repulsion $U$ on the dot. Analogously to the noninteracting case discussed before, from Eq.~(\ref{eq_exp}) we can write the time-dependent probability vector in terms of contributions exhibiting different decay times
\begin{widetext}
\begin{eqnarray}
\label{eq_solution}
\mathbf{P}(t) & = & \mathbf{P}^\mathrm{eq}
+\frac{1}{1-f(\epsilon)+f(\epsilon+U)}\left(\begin{array}{c}
-[1-f(\epsilon)]\\
\frac{1}{2}\left[1-f(\epsilon)-f(\epsilon+U)\right]\\
\frac{1}{2}\left[1-f(\epsilon)-f(\epsilon+U)\right]\\
f(\epsilon+U)
\end{array}\right)e^{-\gamma_n t}\left(\langle\hat{n}\rangle^\mathrm{in}-\langle\hat{n}\rangle^\mathrm{eq}\right)
\nonumber \\
&&
+\left(\begin{array}{c}
0\\
\frac{1}{2}\\
-\frac{1}{2}\\
0
\end{array}\right)
e^{-\gamma_s t}\langle\hat{s}\rangle^\mathrm{in}+\left(
\begin{array}{c}
-1\\1\\1\\-1
\end{array}
\right)
e^{-\gamma_m t}\left(
\langle\hat{m}\rangle^\mathrm{in}-\langle\hat{m}\rangle^\mathrm{eq}
\right)\ .
\end{eqnarray}
\end{widetext}
Again, $\mathbf{P}^\mathrm{eq}$ is the eigenvector of  $\mathbf{A}^{(1)}=\mathbf{W}^{(1)}$ with  the zero eigenvalue and represents the equilibrium state in lowest order in the tunnel coupling (the explicit form of the four-dimensional matrix $\mathbf{A}^{(1)}$, together with its entire set of eigenvalues and eigenvectors, is given in the Appendix).  In the two-particle basis \mbox{$ \{|\chi\rangle\}=\{|0\rangle, |\uparrow\rangle, |\downarrow\rangle, |d\rangle\}$},  again $\hat{n}=\left(0,1,1,2\right)$ represents the charge operator, and $\hat{s}=\left(0,1,-1,0\right)$ represents the spin operator. The form of the operator $\hat{m}$ is modified by the presence of finite Coulomb interaction; the explicit form will be discussed later in this sub-section (see Eq. (\ref{eq_def_m}) below). The initial and equilibrium expectation values for these operators,  entering in the above Eq.~(\ref{eq_solution}), are obtained as $\langle\hat{o}\rangle^{\mathrm{in/eq}}=\hat{o}\cdot\mathbf{P}^\mathrm{in/eq}$, with $\hat{o}=\hat{s},\hat{n},\hat{m}$. Explicit expressions for $\langle\hat{n}\rangle^{\mathrm{eq}}$ and $\langle\hat{m}\rangle^{\mathrm{eq}}$ are shown below.
The negative of the other three eigenvalues of $\mathbf{A}^{(1)}$ directly determine the decay of charge, spin,\cite{splettstoesser10a} and the quantity denoted by $\hat{m}$. These decay rates read
\begin{subequations}
\begin{eqnarray}
\gamma_n&=&\Gamma\left[1+f(\epsilon)-f(\epsilon+U)\right]\label{eq_lcharge1}\\
\gamma_s&=&\Gamma\left[1-f(\epsilon)+f(\epsilon+U)\right]\label{eq_lspin1}\\
\gamma_m&=&2\Gamma.
\label{eq_lmal1}
\end{eqnarray}
\end{subequations}
Notice that due to interaction, the relaxation rates for charge and spin ($\gamma_n$ and $\gamma_s$ respectively) differ from each other and depend on the level position $\epsilon$, in contrast to the non-interacting case. Their dependence on the level position is
shown in Fig.~\ref{fig_decay}.
In the region for $-U<\epsilon<0$, $\gamma_n$ is enhanced as the charge decays into the twofold degenerate state of single-occupation, whereas the spin relaxation in first order in $\Gamma$ is suppressed, since spin-flip processes are not possible. However, the third decay rate, $\gamma_m$, remains fully energy independent as in the case with $U=0$.

The right eigenvectors occurring in Eq.~(\ref{eq_solution}) each represent a change to the steady state density matrix that decays exponentially with rate $\gamma_i$ ($i=n,s,m$). Therefore, a system being brought out of equilibrium by a symmetric deviation between $p_\uparrow$ and $p_\downarrow$ only, is decaying with a rate $\gamma_s$. A deviation from equilibrium in which the occupation of the even sector, $p_0+p_\mathrm{d}$ is symmetrically shifted from the odd sector, $p_\uparrow+p_\downarrow$, is governed solely by the relaxation rate $\gamma_m$. This right eigenvector is found to play an important role also in the low-temperature renormalization of this model.~\cite{saptsov} An energy-dependent change in the occupation probabilities as prescribed by the second vector in Eq.~(\ref{eq_solution}) yields a decay of the total charge of the system with the rate $\gamma_n$. 
The conditions under which specific deviations from the equilibrium state should be performed in order to obtain a specific decay rate, are discussed in the following Section.

The attribution of these relaxation rates to the charge, spin, and $\hat{m}$ arises from the independent decay of these quantities, due to the explicit form of the \textit{left} eigenvectors of $\mathbf{A}^{(1)}$.
The spin operator coincides  with the left eigenvector associated to the eigenvalue $-\gamma_s$ and since it has a vanishing equilibrium value, the time evolution of its expectation value is given by
\begin{equation}\label{eq_srelax}
\begin{array}{cccc}
(0, & 1, & -1, & 0)
\end{array}\cdot
\left(\begin{array}{c}
p_0(t)\\
p_\uparrow(t)\\
p_\downarrow(t)\\
p_\mathrm{d}(t)
\end{array}\right)
= \langle\hat{s}\rangle(t)=e^{-\gamma_s t}\langle\hat{s}\rangle^\mathrm{in}.
\end{equation}

Equivalently, the left eigenvector corresponding to the eigenvalue $-\gamma_n$, is $\hat{n}-\langle\hat{n}\rangle^\mathrm{eq}(1,1,1,1)$. It contains the charge operator $\hat{n}$ and its equilibrium value $\langle\hat{n}\rangle^\mathrm{eq}=2f(\epsilon)/\left[1+f(\epsilon)-f(\epsilon+U)\right]$. Hence, for the time evolution of the charge we find
\begin{eqnarray}\label{eq_nrelax}
\nonumber
\begin{array}{cccc}
(0, & 1, & 1, & 2)
\end{array}
&\cdot&
\left(\begin{array}{c}
p_0(t)\\
p_\uparrow(t)\\
p_\downarrow(t)\\
p_\mathrm{d}(t)
\end{array}\right) -\langle\hat{n}\rangle^\mathrm{eq}\\ \nonumber
&=& \langle\hat{n}\rangle(t) -\langle\hat{n}\rangle^\mathrm{eq}\\
&=&e^{-\gamma_n t}\left(\langle\hat{n}\rangle^\mathrm{in}-\langle\hat{n}\rangle^\mathrm{eq}\right) .
\end{eqnarray}
\begin{figure}[h]
   \includegraphics[width=0.9\linewidth,clip]{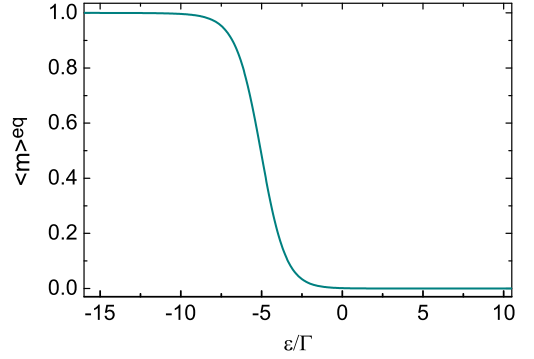}
   \caption{Equilibrium value of the quantity $\hat{m}$ as a function of the dot level position $\epsilon$. The other parameters are: $k_\mathrm{B}T=1.5\Gamma$ and $U=10\Gamma$. }
    \label{fig_meh}
\end{figure}
The quantity decaying with the rate $\gamma_m$ alone is related to the left eigenvector $\hat{m}-\langle\hat{m}\rangle^\mathrm{eq}(1,1,1,1)$, where the operator $\hat{m}$ is given by
\begin{eqnarray}
\hat{m}=\frac{1}{1-f(\epsilon)+f(\epsilon+U)}
\left(
\begin{array}{c}
0\\
f(\epsilon+U)\\
f(\epsilon+U)\\
-1+f(\epsilon)+f(\epsilon+U)
\end{array}\right)^T\ .
\label{eq_def_m}
\end{eqnarray}
Its expectation value follows a time evolution equivalent to the one for the charge in Eq.~(\ref{eq_nrelax}): $\langle\hat{m}\rangle(t) -\langle\hat{m}\rangle^\mathrm{eq}=e^{-\gamma_m t}\left(\langle\hat{m}\rangle^\mathrm{in}-\langle\hat{m}\rangle^\mathrm{eq}\right)$. Its equilibrium value $\langle\hat{m}\rangle^\mathrm{eq}=f(\epsilon)f(\epsilon+U)/\left[1-f(\epsilon)+f(\epsilon+U)\right]$, plotted in Fig.~\ref{fig_meh}, is - in contrast to spin and charge - not sensitive to the regime of single occupation on the quantum dot. Instead, it exhibits a feature close to the electron-hole symmetric point of the Anderson model, indicating that $\hat{m}$ represents a quantity which is affected by two-particle effects and it decays with a rate that is not modified by the Coulomb interaction $U$.

Already for the noninteracting case, we found that the rate $2\Gamma$ appears as a consequence of introducing two particles in the system, and we considered the deviations from equilibrium charge as a quantity involving two-particle processes leading to such a decay rate. Also in the case for finite Coulomb interaction, the time-dependent mean squared deviations $\langle[\hat{n}-\langle \hat{n}\rangle^\mathrm{eq}]^2\rangle(t)$ are suitable to reveal the relaxation rate $\gamma_m=2\Gamma$. Their time evolution is obtained by means of Eq.~(\ref{eq_solution}) and reads
\begin{eqnarray}
\langle[\hat{n}-\langle \hat{n}\rangle^\mathrm{eq}]^2\rangle(t) -
[\langle \hat{n}\rangle^\mathrm{eq}]^2
& = &
C\cdot\langle \hat{n}\rangle(t)-2\cdot\langle
\hat{m}\rangle(t)\nonumber\\
 \label{eq_variance}
\end{eqnarray}
\begin{figure}[h]
   \includegraphics[width=0.9\linewidth,clip]{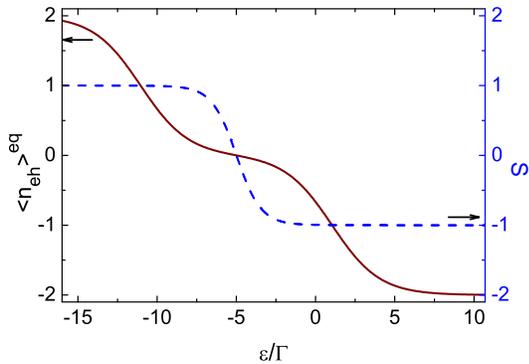}
   \caption{(Color online) Equilibrium electron-hole occupation $\langle \hat{n}_{eh}\rangle^\mathrm{eq}$ (red, solid line) and the coefficient $S$ (blue, dashed line) as a function of the dot level position $\epsilon$. The other parameters are:  $k_\mathrm{B}T=1.5\Gamma$ and $U=10\Gamma$.}
\label{fig_coeff}
\end{figure}
where in front of the time-dependent charge $\langle \hat{n}\rangle(t)$ the following coefficient appears:
\begin{equation}
C=-2\langle p_\mathrm{d}-p_0\rangle^\mathrm{eq}+S \\
\label{ceh}
\end{equation}
with
\begin{equation}
S=-\frac{1-f(\epsilon)-f(\epsilon+U)}{1-f(\epsilon)+f(\epsilon+U)}.\
\label{eq_coeff}
\end{equation}

The quantity $\langle p_\mathrm{d}-p_{0}\rangle^\mathrm{eq}=-\left[1-f(\epsilon)-f(\epsilon+U)\right]/\left[1+f(\epsilon)-f(\epsilon+U)\right]$ is the
difference between the probability of doubly occupied and empty dot in
equilibrium, which can also be related with the occupation of
electrons and holes, $\langle \hat{n}_{eh}\rangle=2\langle p_{\mathrm{d}}-p_{0}\rangle$.
 The behavior of $\langle \hat{n}_{eh}\rangle^\mathrm{eq}$ is shown in  Fig.~\ref{fig_coeff}.
For $\epsilon<-U$, when the dot is doubly occupied,  $\langle \hat{n}_{eh}\rangle^\mathrm{eq}=2$;  for $-U<\epsilon<0$, when one
electron and one hole are present in the system (singly occupied dot), $\langle \hat{n}_{eh}\rangle^\mathrm{eq}=0$; and for $\epsilon>0$, when the system is  completely ``filled with holes'' (empty dot),  $\langle \hat{n}_{eh}\rangle^\mathrm{eq}=-2$.
The quantity $S$ is also shown in Fig.~\ref{fig_coeff} (blue dashed line), exhibiting a sign change around $\epsilon=-U/2$, the point at which the Anderson model is electron-hole symmetric. By replacing
$\epsilon\rightarrow-\epsilon-U$, we go from the electron-like to the hole-like behavior, finding an
inversion in the sign of $S$, $S\rightarrow-S$. The function $S$ therefore indicates whether the spectrum of the quantum dot is electron-like or hole-like.\par

The mean squared deviations of the charge from its value at equilibrium is an example for a physical quantities showing a decay with $\gamma_m$; it also includes the charge relaxation rate $\gamma_n$, which is found independently from the time evolution of the charge. Equivalently also the time-resolved charge variance, $\langle[\hat{n}-\langle \hat{n}\rangle(t)]^2\rangle(t)$, or the time-resolved spin variance,\cite{notacrooker} $\langle(\hat{s})^2\rangle(t)$, contain a contribution decaying with $\gamma_m$.

\subsection{Response to an external perturbation}

We now consider in detail which external perturbations are necessary in order to induce a decay of the \textit{full} occupation probability vector with one certain relaxation rate only, in a controlled way. Furthermore, we address the conditions under which a single decay rate can be extracted more easily from the occupation of a single state by a measurement with a nearby quantum point contact (QPC).

\begin{figure}[h]
 \epsfig{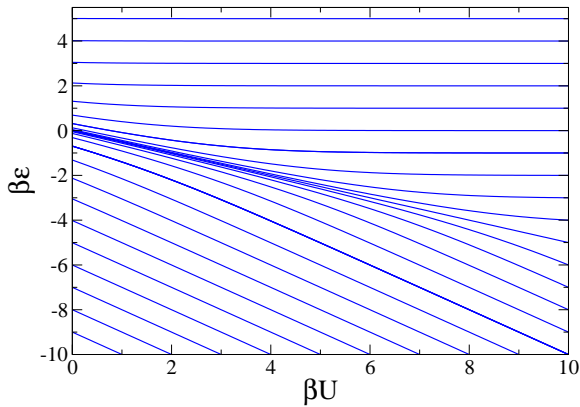}
 \caption{Field lines describing variations of $\epsilon$ and $U$ that lead to a response of the system with only the rate $\gamma_m$.}
    \label{fig_field}
\end{figure}

 We first address the case of an infinitesimal perturbation (linear response). A small variation of the gate potential leads to a decay of the charge governed by the charge relaxation rate $\gamma_n$. Similarly, the infinitesimal variation of the Zeeman splitting in the dot yields a decay with the spin relaxation rate $\gamma_s$. In order to obtain a decay of the state with the rate $\gamma_m$ only, it is not sufficient to modulate the gate voltage, also the two-particle term in the Hamiltonian, $U n_\uparrow n_\downarrow$, needs to be varied.  The on-site repulsion $U$ could be changed, for example, by tuning the carrier density in a nearby two-dimensional electron gas, thereby controlling the screening of the electron-electron interaction in the dot. From Eq.~(\ref{eq_solution}) we know that a dynamics given only by $\gamma_m$ is obtained if the occupation of the even states are changed in the same direction, opposite to that of the  single occupied states; this condition is fulfilled if infinitesimal variations of the gate, $\epsilon\rightarrow\epsilon+d\epsilon$, and of the interaction, $U\rightarrow U+dU$, obey the relation:
\begin{equation}
dU = -\frac{1+2 \exp(\beta\epsilon)+\exp(\beta[2\epsilon+U])}{1+\exp(\beta\epsilon)}d\epsilon.
\label{dvarm}
\end{equation}
This expression is represented in terms of field lines in Fig.~\ref{fig_field}. An infinitesimal change tangential to the field line passing through the point corresponding to the initial values of $\epsilon$ and $U$ leads to a pure decay with $\gamma_m$.

For parameter variations that are not infinitesimal (beyond linear response), a change only of the gate voltage results in a decay of the state with both rates $\gamma_n$ and $\gamma_m$.  From Eq.~(\ref{eq_solution}) we find that a finite variation of the energy level and the interaction from an initial condition $(\epsilon_0,U_0)$ to $(\epsilon=\epsilon_0+\Delta\epsilon,U=U_0+\Delta U)$ resulting in a relaxation containing solely $\gamma_n$, satisfies the equation
\begin{equation}
\beta\Delta U=-\beta\Delta\epsilon+\ln[2-e^{-\beta\Delta\epsilon}].
\label{varn}
\end{equation}
A relaxation given \textit{only} by the rate $\gamma_m$ is found when the relation:
\begin{equation}
U=U_0+\frac{1}{\beta}\ln\left[\frac{e^{\beta(\epsilon_0-\epsilon)}\left(1+e^{\beta\epsilon_0}\right)}{1+e^{\beta\epsilon}+e^{\beta(\epsilon+\epsilon_0+U_0)}-e^{\beta(2\epsilon_0+U_0)}}\right]
\label{varm}
\end{equation}
is fulfilled. For different values of $\epsilon_0$ and $U_0$, Eq.~(\ref{varm}) produces again the field lines shown in Fig.~\ref{fig_field}. Therefore, finite variations of the parameters between two points lying on \textit{the same} field line yield a dynamics for the entire occupation probabilities vector $\mathbf{P}$ governed only by $\gamma_m$. Obviously, a generic variation in both $\epsilon$ and $U$ which does not fulfill the conditions specified by Eqs.~(\ref{varn}) or (\ref{varm}) exhibits a dynamics of the probabilities with two time scales: $\gamma_n$ and $\gamma_m$.

In Fig.~\ref{fig_field} it is observed that in the region $\epsilon>-U/2$ the field lines are approximately  horizontal, i.e, only the interaction $U$ needs to be varied while keeping the level position constant in order to see a dynamics of the probability governed by $\gamma_m$ only. In fact, in this regime the QD is predominantly empty and variations of the interaction strength $U$ do not affect the occupation of the dot. This is the reason why this variation yields a dynamics in which the rate $\gamma_n$ does not contribute. On the other hand, in the region for $\epsilon<-U/2$ in order to avoid that the number of particles on the dot changes, which would lead to a relaxation with rate $\gamma_n$, a variation in $U$ needs to be accompanied by an opposite variation in $\epsilon$, that is $\Delta\epsilon=-\Delta U$. The crossover between the two regimes appears around the symmetry point of the Anderson model, $\epsilon=-U/2$.

Importantly, it is also possible to read out either the rate $\gamma_n$ or the rate $\gamma_m$ by varying the gate voltage only (and, thus, not fulfilling Eqs.~(\ref{dvarm}) and (\ref{varm})), which is easier to realize in an experiment.
This can be done by measuring an observable that is sensitive to only one occupation probability, for instance the probability of the quantum dot being empty.
Such a time-resolved read-out of the probability can be achieved by considering a QPC  located nearby the system and tuned such that it conducts only if the QD is empty.~\cite{field93, vander04,mueller10} In the simplest model of the QPC, which assumes a very fast response, the operator corresponding to the current in the QPC is given by
\begin{equation}
\label{iqp}
\hat{i}_{QPC}=i_0\left(1,0,0,0\right),
\end{equation}
where $i_0$ is a constant current, given by the characteristics of the QPC potential.
The expectation value of the QPC current is simply $\langle\hat{i}_{QPC}\rangle(t)=i_0\langle p_0\rangle(t)$. In this way, the QPC effectively measures the dynamics of the occupation probability $p_0$.
According to Eq.~(\ref{eq_solution}), a modulation of the gate in which the initial value $\langle\hat{m}\rangle^{\mathrm{in}}$ equals the equilibrium value $\langle\hat{m}\rangle^{\mathrm{eq}}$ leads to a pure decay with $\gamma_n$. Instead, for a decay given by $\gamma_m$ either the factor $\langle\hat{n}\rangle^{\mathrm{in}}-\langle\hat{n}\rangle^{\mathrm{eq}}$ or the factor $\left[1-f(\epsilon)\right]/\left[1-f(\epsilon)+f(\epsilon+U)\right]$ in Eq.~(\ref{eq_solution}) has to vanish.
\begin{figure}[h]
   \includegraphics[width=0.9\linewidth,clip]{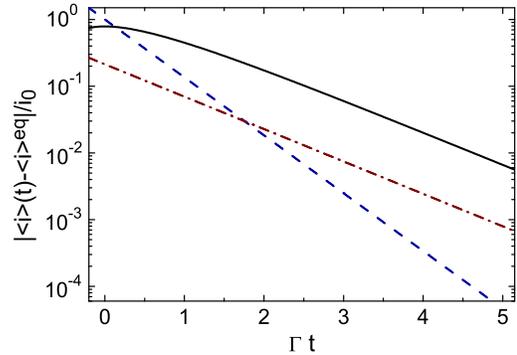}
\caption{(Color online) Logplot of the current in the QPC as a function of the time after a finite variation of $\epsilon$. The on-site Coulomb repulsion $U$ is constant and takes the value $U=5 k_BT$. Dashed blue line: $\epsilon$ changes from $\epsilon_0=10 k_BT$ to $\epsilon=-10 k_BT$, its slope yields the relaxation rate $\gamma_m$.  Red dot-dashed line:  in this case $\epsilon_0 =10 k_BT$ to $\epsilon=2 k_BT$, and the slope leads to $\gamma_n$. The black line is obtained if $\epsilon$ changes from $\epsilon_0=-13 k_BT$ to $\epsilon=2 k_BT$, in which both rates $\gamma_m$ and $\gamma_n$ are present. In all cases we have subtracted the corresponding value for the current in the long-time limit.}
    \label{lines}
\end{figure}

Results for the QPC current for different variations of the level position $\epsilon$ while $U$ is kept constant, are shown in the logarithmic plot in Fig.~\ref{lines}. For clarity, we also subtracted the corresponding current in the long time limit, $\langle\hat{i}\rangle^{\mathrm{eq}}$.
In particular, for a fixed value of $U$ equal to $5k_BT$, we find that if the level position is changed from  $\epsilon_0=10 k_BT$ to $\epsilon=-10 k_BT$, the time evolution of $p_0$ is governed entirely by the rate $\gamma_m$, giving rise to the straight, blue-dashed line in Fig.~\ref{lines}. Its slope is given by $\gamma_m$, making it possible to extract this relaxation rate from measurements of the current in the QPC. However we can  obtain a dynamics of $p_0$ given mainly by the rate $\gamma_n$ by performing a variation in $\epsilon$ from $\epsilon_0=10 k_BT$ to $\epsilon=2 k_BT$ which results in the red dot-dashed straight line in Fig.~\ref{lines}; again, the slope yields the corresponding relaxation rate which takes the value $\gamma_n=1.12~\Gamma$. Finally, we show an example in which variations from $\epsilon_0=-13 k_BT$ to $\epsilon=2 k_BT$ (solid black line) produce a dynamics of $p_0$ which includes two exponential decays with rates $\gamma_m$ and $\gamma_n$.  As a result, the curve exhibits a change in the slope, showing that a single rate will not be obtained by arbitrary variations of the parameters.

\subsection{Second-order corrections  in the tunnel coupling}

In the previous sections we investigated the relaxation rates in first order in the tunnel coupling strength $\Gamma$. However, corrections due to higher order tunneling processes appear when the tunnel coupling gets stronger. Besides quantitative corrections, this reveals an interesting new aspect.  In second order in the tunnel coupling, the matrix $\mathbf{A}^{(2)}$ included in the exponential decay takes the form $\mathbf{A}^{(2)}=\mathbf{W}^{(2)}+\partial \mathbf{W}^{(1)}\cdot \mathbf{W}^{(1)}$. The second-order corrections to the  relaxation rates for charge and spin are given by:\cite{splettstoesser10a}
\begin{eqnarray}
\gamma_\mathrm{n}^{(2)} & = &  \sigma(\epsilon,\Gamma,U)\frac{\partial}{\partial\epsilon}\gamma_n+ \sigma_\Gamma(\epsilon,\Gamma,U)\gamma_\mathrm{n}\nonumber\\
 & &+2 \frac{
                 f(\epsilon+U)W_{0\mathrm{d}}
                 +
                  \left[1-f(\epsilon)\right]W_{\mathrm{d}0}
                  }
                   {1-f(\epsilon)+f(\epsilon+U)} \label{eq_charge_corr}\\
\gamma_\mathrm{s}^{(2)} & = & \sigma(\epsilon,\Gamma,U) \frac{\partial}{\partial\epsilon}\gamma_\mathrm{s}+\sigma_\Gamma(\epsilon,\Gamma,U)\gamma_\mathrm{s}+2W_\mathrm{sf}\ . \label{eq_spin_corr}
\end{eqnarray}
These corrections contain renormalization terms as well as real cotunneling contributions. On one hand, the renormalization terms contain an effect due to the level renormalization $\epsilon\rightarrow\epsilon+\sigma(\epsilon,\Gamma,U)$, with $\sigma(\epsilon,\Gamma,U)=\Gamma[\phi(\epsilon+U)-\phi(\epsilon)]$, $\phi(\epsilon)=\frac{1}{2\pi}\mathrm{Re}\Psi\left(\frac{1}{2}+i\frac{\beta\epsilon}{2\pi}\right)$ and $\Psi(x)$ is the digamma function. On the other hand, the renormalization of the tunnel coupling appears, $\Gamma\rightarrow\Gamma[1+\sigma_\Gamma(\epsilon,\Gamma,U)]$, with $\sigma_\Gamma(\epsilon,\Gamma,U)= -S\left[\Gamma\phi'(\epsilon)+\Gamma\phi'(\epsilon+U)-\frac{2}{U}\sigma(\epsilon,\Gamma,U)\right]$ and where $S$ was defined in Eq.~(\ref{eq_coeff}). Real cotunneling contributions are manifest in terms of spin flips, $W_\mathrm{sf}$, and coherent transitions changing the particle number on the dot by $2$, $W_{0d}$ and $W_{d0}$. These cotunneling terms read
\begin{eqnarray}
W_\mathrm{sf} & = & -\frac{\Gamma}{\beta}\left[\Gamma\phi''(\epsilon)+\Gamma\phi''(\epsilon+U)-\frac{2}{U}\sigma'(\epsilon,U)\right]\\
W_{\mathrm{d}0} & = & - \frac{2 \Gamma} {e^{\beta(2\epsilon+U)}-1}  \left[\Gamma\phi'(\epsilon)+\Gamma\phi'(\epsilon+U)-\frac{2}{U}\sigma(\epsilon,U)\right],\nonumber\\
\end{eqnarray}
and $W_{0\mathrm{d}} =  \exp[\beta(2\epsilon+U)] W_{\mathrm{d}0}$.

The way in which the cotunneling contributions enter in the respective charge and spin relaxation rates is related to the deviation of the state of the QD from equilibrium, given by Eq.~(\ref{eq_solution}) in first order in $\Gamma$.  As an example we discuss the correction to the charge decay rate, second line of Eq.~(\ref{eq_charge_corr}). There the factor $2$ appears due to the change in the charge by $\pm2$ in a process bringing the dot from zero to double occupation and vice versa.~\cite{leijnse09} The fraction with which the transition from zero to double occupation, $W_{d0}$,  enters the correction to the charge relaxation rate, $\gamma_n^{(2)}$, is given by the deviation from equilibrium of $\mathbf{P}(t)$ in the direction of $p_0$, of the contribution which actually decays with $\gamma_n$ only. This is the first component of the second vector in Eq.~(\ref{eq_solution}). Equivalently, the transition from double to zero occupation, $W_{0d}$, enters with the fraction given by the fourth component of the same vector, namely by the deviation from equilibrium of $\mathbf{P}(t)$ in the direction of $p_d$.

Strikingly, in contrast to the charge and spin relaxation rates, $\gamma_m$  does not get renormalized at all by second order tunneling processes:
\begin{eqnarray}
\gamma_m^{(2)} & = & \gamma_m^{(1)}\sigma_\Gamma(\epsilon,\Gamma,U)+\frac{1-f(\epsilon)-f(\epsilon+U)}{1-f(\epsilon)+f(\epsilon+U)}\left(W_{0d}-W_{d0}\right) \nonumber\\
\label{m2ndorder}
& = & 0\ ,
\end{eqnarray}
The reason for this is that the contribution due to $\Gamma$ renormalization and those due to coherent processes between empty and doubly occupied dot, cancel each other. The lack of second order corrections, confirms that this relaxation rate is related to a quantity which is not sensitive to the Coulomb interaction. The fact that corrections are missing, is also found using a renormalization-group approach.~\cite{saptsov}.

Another important aspect of this missing second-order correction is that it is due to an exact cancelation of the contribution due to virtual second order processes, namely the $\Gamma$-renormalization, with real cotunneling contributions. This is in contrast to, e.g. the conductance, where only the real cotunneling processes contribute far from resonances, while renormalization terms are limited to the resonant regions.

\subsection{Hybrid systems}

Until now, we considered the quantum dot to be coupled to a normal conducting lead. However, the vicinity of a superconducting or a ferromagnetic reservoir induces correlations between electrons and holes or between charge and spin, respectively.
 In the following we study, in first order in the tunnel coupling strength $\Gamma$,  the influence of induced correlations on the relaxation rates of   the dot.
The charge response of a noninteracting mesoscopic scattering region coupled to both normal and superconducting leads has been studied in Refs.~\onlinecite{Pilgram02,xing07}.

\subsubsection{Proximity to a superconducting lead}

In the previous sections we have seen that the rate $\gamma_m$, which together with the time decay of charge and spin determines the relaxation of the QD to the equilibrium state, is independent of the level position and the Coulomb interaction and that it enters in the time evolution of quantities sensitive to two-particle effects.
It is therefore expected that the rate $\gamma_m$ will directly influence the relaxation of the charge towards the equilibrium in a setup that naturally mixes the empty and doubly occupied states of the dot. This situation is obtained if the QD is not only coupled to a normal lead (with tunnel coupling strength $\Gamma$) but also to an \textit{additional} superconducting contact (with tunnel coupling strength $\Gamma_S$), as shown in Fig.~\ref{fig_scheme} (b). We consider only the case when the superconductor is kept at the same chemical potential as the normal lead and we set both chemical potentials to zero.
The only purpose of the extra lead is here to induce superconducting correlations on the dot via the proximity effect.
To the original Hamiltonian, $H_D+H_\mathrm{res}+H_T$, we now add the Hamiltonian for the superconducting contact and its tunnel coupling to the QD,
\begin{eqnarray}
H_\mathrm{S} & = &
\sum_{k\sigma}\epsilon_{\mathrm{S} k}c^\dagger_{\mathrm{S} k\sigma}c^{}_{\mathrm{S} k\sigma}-\sum_k\left(\Delta c^{}_{\mathrm{S}-k\downarrow}c^{}_{\mathrm{S} k\uparrow}+\mathrm{H.c.}\right)\nonumber\\
&&+\sum_{k,\sigma}(V_\mathrm{S}
c_{\mathrm{S}k\sigma}^{\dagger}
d_{\sigma}+\mathrm{H.c})\, .
\end{eqnarray}
where  $c^{(\dagger)}_{\mathrm{S} k\sigma}$ is the the annihilation (creation) operator of electrons in the lead.
In the limit of a large superconducting gap $\Delta$ the effect of the additional contact can be cast in an effective Hamiltonian of the dot which includes a coupling between electrons and holes in the QD, $H_{D}^\mathrm{(eff)}=H_D
-\Gamma_s/2(d_{\downarrow}^{\dagger}d_\uparrow^{\dagger}+ \mathrm{H.c})$.
The eigenstates of the proximized dot are the states of single occupation $|\sigma\rangle$ and other two states which are superpositions of the empty and double occupied states of the dot (due to Andreev reflection):
\begin{equation}
|\pm\rangle=\frac{1}{\sqrt{2}}\sqrt{1\mp\frac{\delta}{2\epsilon_A}}|0\rangle\mp\frac{1}{\sqrt{2}}\sqrt{1\pm\frac{\delta}{2\epsilon_A}}|d\rangle
\label{eq_pmvectors}
\end{equation}
with energies given by $E_{\pm}=\delta/2\pm\epsilon_A$, where the level
detuning between $|0\rangle$ and $|d\rangle$ is $\delta=2\epsilon+U$ and $2\epsilon_A=\sqrt{\delta^2+\Gamma_{s}^{2}}$ is the energy  splitting between the $|+\rangle$ and $|-\rangle$ states.\cite{braggio11,eldridge10} In the new basis $\left\{|+\rangle,|\uparrow\rangle,|\downarrow\rangle,|-\rangle\right\}$, the vector representing the charge operator is expressed as $\hat{n}=\left(-\sqrt{2+\frac{\delta}{\epsilon_A}},1,1,\sqrt{2-\frac{\delta}{\epsilon_A}}\right)$ and we expect that the effect of the mixing of electrons and holes will be visible in its time evolution. In first order in the tunnel-coupling strength to the normal reservoir $\Gamma$ and assuming $\Gamma\ll\Gamma_{S}$, we find the relaxation rates
\begin{eqnarray}
\gamma_{S,1}&=&\Gamma\left[1+f\left(\epsilon-E_-\right)-f\left(E_+-\epsilon\right)\right]\\
\gamma_{S,s}&=& \Gamma\left[1-f(\epsilon-E_-)+f(E_+-\epsilon)\right]\\
\gamma_{S,2}&=&2\Gamma.
\end{eqnarray}

Remarkably the eigenvalue $-2\Gamma=-\gamma_{S,2}$ remains unaffected, i.e. $\gamma_m=\gamma_{S,2}$ is not modified by the presence of the additional superconducting lead.

The spin on the dot, which is determined by the occupation probabilities of singly occupied states, still decays with a single relaxation rate given by $\gamma_{S,s}$, i.e.  $\mathbf{l}_{s}=\left(0,1,-1,0\right)$ is an eigenvector of the kernel $\mathbf{A}^{(1)}$ (in the proximized basis).
In contrast, the decay of the charge to its equilibrium value is  given by
\begin{eqnarray}
\langle\hat{n}\rangle_{SC}(t) & = & \frac{1}{2}
                \left[
                       \langle \hat{n}\rangle^{\mathrm{in}}-\langle \hat{n}\rangle^{\mathrm{eq}}
                       \right]
                  \left(e^{-\gamma_{S,2} t}+e^{-\gamma_{S,1}t}
                        \right)
                        +\langle \hat{n}\rangle^{\mathrm{eq}}\nonumber\\
& & +a_{SC}
      \frac{1}{2}\left[\langle x\rangle^{\mathrm{in}}-\langle x\rangle^{\mathrm{eq}}\right]
      \left(e^{-\gamma_{S,2} t}-e^{-\gamma_{S,1}t}\right) \nonumber \\
& & +\frac{1}{2}\left[\langle y\rangle^{\mathrm{in}}-\langle y\rangle^{\mathrm{eq}}\right]
      \left(e^{-\gamma_{S,2} t}-e^{-\gamma_{S,1}t}\right)
\label{eq_nsct}
\end{eqnarray}
with
\begin{eqnarray}
a_{SC} & = & (2-k_--k_+)\frac{f\left(E_+-\epsilon\right)-f\left(E_--\epsilon\right)}{f\left(E_+-\epsilon\right)+f\left(E_--\epsilon\right)} \nonumber
\label{factors_nsct}
\end{eqnarray}
and where we defined the difference in the occupation of the $|\pm\rangle$ states, $x=p_+-p_-$ and the quantity $y=(k_--1)p_++(k_+-1)p_-$, with $k_{\pm}=\mp\sqrt{2\pm\frac{\delta}{\epsilon_A}}$.
The charge
evolves  with two different time scales,
$\gamma_{S,1}$ and $\gamma_{S,2}=\gamma_m$,
instead of only one as in the normal case. This is a direct consequence of the mixing of the states $|0\rangle$ and $|d\rangle$ induced by the superconducting contact. This effect opens the possibility to extract this rate by measuring the time evolution of the charge in the proximized dot.

\subsubsection{Ferromagnetic lead}

Even though the presence of a superconducting lead couples electrons and holes, the relaxation rate $\gamma_m$ has not been modified. Since we associate this rate with processes involving two particles each with spin $\sigma$, it is expected that if the spin symmetry is broken by introducing a ferromagnetic contact, the rate $\gamma_m$ will now be the sum of the tunneling rates for spin up and spin down electrons.
In order to verify this, we consider the Hamiltonian used for the normal case and assume a spin-dependent density of states in the only reservoir attached to the quantum dot, see Fig.~\ref{fig_scheme} (c). This leads to spin-dependent tunnel couplings, $\Gamma_\uparrow$ and $\Gamma_\downarrow$, which are included in the corresponding transition
matrix $\mathbf{A}^{(1)}$. Diagonalization of $\mathbf{A}^{(1)}$ yields the three relaxation rates:
\begin{eqnarray}
\gamma_{F,1} & = & \Gamma +\frac{1}{2}\sqrt{\left(\Delta\Gamma\right)^2+4\Gamma_\uparrow\Gamma_\downarrow\left[f(\epsilon)-f(\epsilon+U)\right]^2} \\
\gamma_{F,2} & = & \Gamma-\frac{1}{2}\sqrt{\left(\Delta\Gamma\right)^2+4\Gamma_\uparrow\Gamma_\downarrow\left[f(\epsilon)-f(\epsilon+U)\right]^2} \\
\gamma_{F,m}&=& 2 \Gamma
\end{eqnarray}
with $\Gamma=\frac{1}{2} \left(\Gamma_\downarrow+\Gamma_\uparrow\right) $ and $\Delta\Gamma=\Gamma_\uparrow-\Gamma_\downarrow$.

As in the normal case, there is an eigenvalue which does not depend on the level position nor on the interaction but on the sum of the different tunneling rates:  $-2\Gamma\rightarrow-\left(\Gamma^{\uparrow}+\Gamma^{\downarrow}\right)$. The appearance of such a combination of the spin-dependent tunneling strengths in the relaxation rate, confirms the statement that two-particle processes involving electrons with both spin polarizations are at the basis of the decay rate $\gamma_m$.

Due to the ferromagnetic lead, the dynamics of spin and charge are now mixed.
The corresponding time evolution in first order in the tunnel coupling takes the form:
\begin{eqnarray}
\langle \hat{s}\rangle_F(t) & = & \frac{1}{2}\langle \hat{s}\rangle^{\mathrm{in}}(e^{-\gamma_{F,1}t}+e^{-\gamma_{F,2}t})\nonumber\\
& & +a_s\langle \hat{s}\rangle^{\mathrm{in}}(e^{-\gamma_{F,1}t}-e^{-\gamma_{F,2}t}) \label{spin_ferro}\\
& & +b_s\left[\langle \hat{n}\rangle^{\mathrm{in}}-\langle \hat{n}\rangle^{\mathrm{eq}}\right](e^{-\gamma_{F,1}t}-e^{-\gamma_{F,2}t}) \nonumber \\
\langle\hat{n}\rangle_F(t) & = & \frac{1}{2}\left[\langle \hat{n}\rangle^{\mathrm{in}}-\langle \hat{n}\rangle^{\mathrm{eq}}\right](e^{-\gamma_{F,1}t}+e^{-\gamma_{F,2}t}) +\langle \hat{n}\rangle^{\mathrm{eq}}\nonumber \\
& & +a_c\left[\langle \hat{n}\rangle^{\mathrm{in}}-\langle \hat{n}\rangle^{\mathrm{eq}}\right](e^{-\gamma_{F,1}t}-e^{-\gamma_{F,2}t})\nonumber \\
& & +b_c\langle \hat{s}\rangle^{\mathrm{in}}(e^{-\gamma_{F,1}t}-e^{-\gamma_{F,2}t})\label{eq_decay_ferro}
\end{eqnarray}
where we introduced the abbreviations:
\begin{eqnarray}
a_s & = & \frac{\Gamma[f(\epsilon)-f(\epsilon+U)]}{2\sqrt{\Delta\Gamma^2+4\Gamma_\uparrow\Gamma_\downarrow\left[f(\epsilon)-f(\epsilon+U)\right]^2}} \nonumber \\
b_s & = &\frac{\Delta\Gamma[1+f(\epsilon)-f(\epsilon+U)]}{2\sqrt{\Delta\Gamma^2+4\Gamma_\uparrow\Gamma_\downarrow\left[f(\epsilon)-f(\epsilon+U)\right]^2}} \nonumber \\
a_c & = & a_s\nonumber \\
b_c & = & \frac{\Delta\Gamma[1-f(\epsilon)+f(\epsilon+U)]}{2\sqrt{\Delta\Gamma^2+4\Gamma_\uparrow\Gamma_\downarrow\left[f(\epsilon)-f(\epsilon+U)\right]^2}}. \nonumber
\end{eqnarray}
The last term in Eq.~(\ref{spin_ferro}) shows that at finite time $t$ the initial charge influences the time evolution of the spin; similarly, the initial spin enters explicitly in the dynamics of the charge, Eq.~(\ref{eq_decay_ferro}). These terms persist in the non-interacting limit, revealing that the coupled evolution of charge and spin including two relaxation rates (which for the non-interacting case take the form $\gamma_{F,1}=\Gamma_\uparrow$ and $\gamma_{F,2}=\Gamma_\downarrow$) is a direct consequence of the presence of the ferromagnetic contact. In contrast, the factor $a_s=a_c$ vanishes for $U=0$ implying that it stems from the combined effect of the Coulomb interaction and the breaking of the spin symmetry.
As expected the independent evolution of charge and spin is recovered in the limit $\Gamma_\uparrow=\Gamma_\downarrow$.
The mixing of the dynamics of both, charge and spin, induced here by a ferromagnetic lead was found in Ref. \onlinecite{splettstoesser10a} for the case of lifted spin-degeneracy in the dot due to a finite Zeeman splitting. Note that for the hybrid as well as for the normal system, the sum of the energy-dependent relaxation rates equals $2\Gamma$, as long as the tunnel coupling $\Gamma$ is treated in first order, only.

\section{Conclusion}

We have studied the different time scales present in the evolution of the reduced density matrix of a
single-level QD with Coulomb interaction and tunnel coupled to a single
reservoir, after being brought out of equilibrium.
Besides the relaxation rates for charge and spin, we find an additional rate $\gamma_m=2\Gamma$, which is
independent of the energy level of the dot as well as of the interaction strength. This relaxation is related to the presence of two particles in the dot and is found to be not sensitive to the Coulomb interaction. The time evolution of the square deviations of the charge from its equilibrium
value is proposed as a physical quantity related with processes involving two-particles leading to the rate $2\Gamma$.
In order to further elucidate the properties of this decay, we analyzed the response of the system to specific variations of both, the interaction strength $U$ and the level position $\epsilon$, finding that $\gamma_m$ can be extracted from time-resolved measurements of the current passing through a nearby quantum point contact.
Additionally, we analyzed two other setups:
a dot proximized by a superconductor and coupled to a normal reservoir, and a dot coupled to a ferromagnetic lead. In the hybrid normal-superconducting systems, we found that the time-resolved read-out of the charge represents another possibility to get access to the rate $\gamma_m$.

\begin{acknowledgments}
We thank Michael Moskalets, Roman Riwar and Maarten Wegewijs for fruitful discussion. Financial support by the Ministry of Innovation, NRW, the DFG via SPP 1285 and KO 1987/5, the European Community's Seventh Framework Programme under Grant Agreement No. 238345 (GEOMDISS), as well as the Swiss National Science Foundation, the Swiss centers of excellence MaNEP and QSIT and the European Marie Curie ITN, NanoCTM is acknowledged.
\end{acknowledgments}

\appendix
\section*{\label{app_vectors} Appendix: Normal case. Eigenvalues and eigenvectors in first order in the tunnel coupling}

The transition matrix for the normal case in the eigenbasis of the isolated QD $\{|0\rangle,|\uparrow\rangle,|\downarrow\rangle,|d\rangle\}$, in first order in the tunneling strength $\Gamma$, is calculated by means of Fermi's Golden rule and is given by:

\begin{widetext}
\begin{eqnarray}
\mathbf{A}^{(1)}=\mathbf{W}^{(1)}(z=0) & = &\Gamma\left(
\begin{array}{cccc}
-2f(\epsilon) & 1-f(\epsilon) & 1-f(\epsilon) & 0\\
f(\epsilon) & -\left[1-f(\epsilon)+f(\epsilon+U)\right] & 0 & 1-f(\epsilon+U))\\
f(\epsilon) & 0& -\left[1-f(\epsilon)+f(\epsilon+U)\right] & 1-f(\epsilon+U))\\
0 & f(\epsilon+U) & f(\epsilon+U) & -2\left[1-f(\epsilon+U)\right]
\end{array}
\right)
\end{eqnarray}
with the Fermi function $f(x)=1/[1+\exp(\beta x)]$, where $\beta$ is the inverse temperature $\beta=(k_BT)^{-1}$.

As $\mathbf{A}^{(1)}$ is non-Hermitian it has different right and left eigenvectors, $\mathbf{r}_i$ and $\mathbf{l}_i$. For a system with a well-defined steady state (as the one we are considering here) there must be at least a zero eigenvalue,  $\lambda_0=0$. \cite{jakob04} The other eigenvalues are found to be the negative of
\begin{eqnarray}
\gamma_n&=&\Gamma\left[1+f(\epsilon)-f(\epsilon+U)\right] \nonumber \\
\gamma_s&=&\Gamma\left[1-f(\epsilon)+f(\epsilon+U)\right] \\
\gamma_m&=&2\Gamma. \nonumber
\label{evalues}
\end{eqnarray}
The right eigenvector corresponding with the zero eigenvalue, $\mathbf{r}_0$, determines the stationary density matrix (which we also label as $\mathbf{P}^{\mathrm{eq}}$), whereas each one of the rest of the right eigenvectors represents a deviation out of the equilibrium density matrix which decays exponentially with a rate given by the negative of the corresponding eigenvalue:

\begin{eqnarray}
\nonumber
\mathbf{r}_0=\frac{1}{1+f(\epsilon)-f(\epsilon+U)}\left(\begin{array}{c}
[1-f(\epsilon)][1-f(\epsilon+U)]\\
f(\epsilon)[1-f(\epsilon+U)]\\
f(\epsilon)[1-f(\epsilon+U)]\\
f(\epsilon)f(\epsilon+U)\\
\end{array}\right), \ \mathbf{r}_s=\frac{1}{2}\left(\begin{array}{c}0\\1 \\-1 \\ 0\end{array}\right),\\
\mathbf{r}_n   = \frac{1}{1-f(\epsilon)+f(\epsilon+U)}\left(\begin{array}{c}
-[1-f(\epsilon)]\\
\frac{1}{2}\left[1-f(\epsilon)-f(\epsilon+U)\right]\\
\frac{1}{2}\left[1-f(\epsilon)-f(\epsilon+U)\right]\\
f(\epsilon+U)
\end{array}\right), \
\mathbf{r}_m   = \left(\begin{array}{c} -1\\1\\1\\-1\end{array}\right)\ .
\end{eqnarray}
The vector of equilibrium occupations $\mathbf{r}_0$ can equivalently be written in terms of the Gibbs factors as
\begin{equation}
\mathbf{r}_0  = \mathbf{P}^{\mathrm{eq}}= \frac{1}{1+2e^{-\beta\epsilon}+e^{-\beta(2\epsilon+U)}}\left(\begin{array}{c}1\\e^{-\beta\epsilon}\\e^{-\beta\epsilon}\\e^{-\beta(2\epsilon+U)} \end{array}\right).
\end{equation}

The left eigenvectors determine the quantities decaying into equilibrium with a single time scale only, and are found to be
\begin{subequations}
\begin{eqnarray}
\mathbf{l}_0 & = & \left(1,1,1,1\right)\\
\mathbf{l}_s & = & \left(0,1,-1,0\right)\\
\mathbf{l}_n & = & \langle \hat{n}\rangle^\mathrm{eq}\left(1,1,1,1\right)-\left(0,1,1,2\right)\\
\mathbf{l}_m & = & \langle \hat{m}\rangle^\mathrm{eq}\left(1,1,1,1\right)-\frac{1}{1-f(\epsilon)+f(\epsilon+U)}\left(0,f(\epsilon+U),f(\epsilon+U),-1+f(\epsilon)+f(\epsilon+U)\right)\ .
\end{eqnarray}
\end{subequations}
\end{widetext}
where $\langle \hat{n}\rangle^\mathrm{eq}=2f(\epsilon)/[1+f(\epsilon)-f(\epsilon+U)]$ and $\langle \hat{m}\rangle^\mathrm{eq}=[f(\epsilon)f(\epsilon+U)]/[1-f(\epsilon)+f(\epsilon+U)]$.

Normalization constants are chosen such that the eigenvectors fulfill the relation $\mathbf{l}_i\cdot\mathbf{r}_j=\delta_{ij}$,\cite{jakob04} with $i,j\in\left\{0,s,n,m\right\}$.

These left eigenvectors contain the operators for spin, charge and $\hat{m}$ in vector representation, which can be understood in the following manner. While in general the expectation value of an operator $\hat{O}$ is found from $\langle\hat{O}\rangle(t)=\mathrm{Tr}\left\{\hat{O}\rho(t)\right\}$, with the full density matrix $\rho$, this can be considerably simplified in the situation considered here, where only diagonal elements of the reduced density matrix of the quantum dot, collected in the vector $\mathbf{P}$, play a role. The expectation value of a quantum dot operator is then obtained by multiplying its vector representation from the left hand side onto the vector $\mathbf{P}$. To show an example the expectation value of the spin on the dot is obtained by multiplying $\mathbf{P}$ from left by  the vector $(0,1,-1,0)$, yielding $\langle\hat{s}\rangle=1\cdot p_\uparrow+ (-1)\cdot p_\downarrow$.
Similarly, all other operators for quantum dot observables can be expressed in such a vector representation.



\end{document}